\documentclass{aa}  
\AddToHook{begindocument/before}{\usepackage{hyperref}}
\usepackage{graphicx}
\usepackage{txfonts}

\begin{document}

   \title{The JWST view of the barred galaxy population in the SMACS0723 galaxy cluster}


   \author{J. M\'endez-Abreu\inst{1,2}, L. Costantin\inst{3},
          \and
          S. Kruk\inst{4}
          }
   \institute{Instituto de Astrof\'isica de Canarias, Calle V\'ia L\'actea s/n, E-38200, La Laguna, Tenerife, Spain\\
              \email{jairomendezabreu@gmail.com}
         \and
             Departamento de Astrof\'isica, Universidad de La Laguna, E-38205, La Laguna, Tenerife, Spain
             \and
             Centro de Astrobiolog\'ia (CAB) INTA-CSIC, Ctra de Ajalvir km 4, Torrej\'on de Ardoz, 28850, Madrid, Spain
             \and
             European Space Agency (ESA), European Space Astronomy Centre (ESAC), Camino Bajo del Castillo s/n, 28692, Villaneuva de la Ca\~{n}ada, Madrid
             }

   \date{Received September 15, 1996; accepted March 16, 1997}

 
  \abstract
   {The cosmic evolution of the barred galaxy population provides key information about the secular evolution of galaxies and the settling of rotationally dominated discs.}
   {We study the bar fraction in the SMACSJ0723.37323 (SMACS0723) cluster of galaxies at $z = 0.39$ using the Early Release Observations obtained with the NIRCam instrument mounted on the JWST telescope.}
   {We visually inspected all cluster member galaxies using the images from the NIRCam F200W filter. We classified the galaxies into ellipticals and discs and determine the presence of a bar. The cluster member selection was based on a combined method using both the available spectroscopy and the color-magnitude relation.}
   {As has previously been found in nearby galaxy samples, we find that the bar fraction distribution of SMACS0723 is a strong function of the galaxy stellar luminosity (or stellar mass). The analogy with local clusters, such as Virgo and Coma, reveals a similar distribution among the three clusters for low-mass galaxies ($\log(M_{\star}/M_{\sun}) \leq$ 9.5). The comparison with a sample of local galaxies in a field environment shows a remarkable lack of bars in this low-mass regime for the SMACS0723 cluster (and, therefore, in Virgo and Coma) with respect to the field. At high masses ($\log(M_{\star}/M_{\sun}) \geq$ 10.25), galaxies in SMACS0723 show a slightly lower bar fraction than those in Coma. At these high masses, we find a much larger bar fraction in SMACS0723 than previous works on field galaxies at $z\sim0.4$. Nevertheless, the difference is only marginal when we compare with a sample of well-resolved local field galaxies. Thus, we suggest that the improved capabilities of JWST with respect to HST in terms of both spatial resolution and image depth are responsible for the  higher bar fraction we obtained.} 
   {Our results support a scenario where cluster environment affects the formation of bars in a mass-dependent way. At high masses, the mild increase in the bar fraction of local clusters (Coma) with respect to both SMACS0723 and local field galaxies suggests a weak effect coming from the cluster environment possibly triggering bar formation. On the other hand, low-mass galaxies show the same bar fraction in the three clusters (different redshifts) and a significant drop with respect to field galaxies at $z=0$, thus suggesting that: i) the bar fraction of low-mass galaxies in clusters is not evolving during the last 4~Gyr; and ii) bar formation is severely inhibited in low-mass galaxies residing in clusters.}


   \maketitle
%

\section{Introduction}
The central role of stellar bars in the secular evolution of disc galaxies is widely accepted. They represent the main structure modifying the morphology of galaxies in the central $\sim$10 kpc \citep{hubble26, buta15} and influence the angular momentum redistribution between the baryonic and dark matter components of the galaxy \citep{debattistasellwood00,martinezvalpuesta07,sellwood14}. Moreover, they have the ability to funnel material towards the galaxy center where starbursts can ignite \citep{martinetfriedly97,sheth05,ellison11}, contributing to the formation
of bulge-like structures \citep{kormendykennicutt04,athanassoula05, bittner20,gadotti20}, inner star-forming rings \citep{buta95,munoztunon04}, and inner bars \citep{erwin04,delorenzocaceres19a,mendezabreu19c,delorenzocaceres20}, thus feeding the central black hole \citep{shlosman90}.

The importance of bars in shaping our understanding of galaxy evolution is also supported by their ubiquity in disc galaxies in the local Universe ($z<0.1$). The general consensus indicates that bars are present in $\sim$50\% of disc galaxies if observed at optical wavelengths \citep{aguerri09,barazza08} and this fraction is slightly increased when using infrared images \citep{eskridge00,marinovajogee07,menendezdelmestre07,erwin18}. Nevertheless, large differences on the bar fraction are still found when analysing different samples. Some authors claim that these differences can be accounted for once the mass dependence of the bar fraction is taken into account \citep{mendezabreu10a, mendezabreu12}, others refer to the gas fraction as the culprit of these variations \citep{masters11,skibba12}; also, the effect of spatial resolution in detecting the smallest bars might also have some influence \citep{erwin18}. Numerical simulations predict that bars spontaneously form due to instabilities in dynamically cold discs \citep{ostrikerpeebles73}, so the answer to the question of why not all local spirals have a bar is still unclear.

The role of the environment in triggering the formation of bars has been a matter of discussion for a long time. \citet{thompson81} claimed that the bar fraction of Coma galaxies increases toward the core of the cluster. Similar results were found for the Virgo and Fornax Clusters \citep{andersen96,eskridge00} and for clusters at intermediate redshifts \citep{barazza09}. In addition, observations seem to favor an increase of the bar fraction in galaxy pairs \citep{kumai86, elmegreen90,giuricin93,varela04}. Tidal interactions in galaxy pairs have been suggested to induce off-center bars in low-mass galaxies \citep{pardy16}, but the observational evidence is still inconclusive \citep{kruk17}. On the other hand, according to \citet{vandenbergh02}, \citet{aguerri09}, and \citet{li09} the bar fraction strongly depends on the properties of the host galaxies but not on their environment. Additionally, \citet{lee12}  claimed that the bar fraction does not depend on the environment when color and central velocity dispersion are fixed. \citet{martinezmuriel11} found that the bar population does not significantly depend on either group mass or on the distance to the nearest neighbour. \citet{giordano11} compared two carefully selected samples that are representative of isolated and cluster galaxies, whereas \citet{marinova12} investigated the bar fraction in lenticular galaxies across different environments which span two orders of magnitude in galaxy density. Neither of them found significant differences. In \citet{mendezabreu12}, they found that the effect of the environment on the bar formation depends on the mass of the galaxy. They proposed that interactions trigger bar formation in massive galaxies, which are stable enough to keep their cold discs even in galaxy clusters. In contrast, the discs of low-mass galaxies are heated by interactions inhibiting the bar formation. 

Numerical simulations have also addressed the influence of the environment in the formation of bars. In addition to the spontaneous bar formation occurring during the secular evolution of galaxies, interactions with other galaxies represent another path to the formation of bars \citep{noguchi87, lokas18}. These tidally induced bars might be the result of a minor merger \citep{gerin90} or a fly-by interaction \citep{martinezvalpuesta17,peschkenlokas19}. Bars formed by interaction-driven mechanisms present a different evolution with respect to those that are spontaneously formed, and their properties will depend on several internal (mass surface density, stellar velocity dispersion, gas fraction) and external (mass of the perturber, impact orbit) properties.

Still, most of our knowledge about the formation and evolution of bars has been produced using local galaxy samples. The Universe at high redshift ($z>$1) was much more violent and turbulent than nowadays. Thus, since a dynamically cold disc is a necessary condition to form a bar, the evolution of the bar fraction is directly related to the evolution of discs. Previous studies carried out using the Hubble Space Telescope (HST) have shown a mixed bag of results regarding the bar fraction evolution with redshift. \citet{abraham99} and \citet{vandenbergh02} argued in favour of a decreasing bar fraction with increasing redshift up to $z \sim$1. Later, the analyses carried out by \citet{elmegreen04} and \citet{jogee04} measured a constant bar fraction up to $z\sim 1$. The situation moves back to a decrease with redshift when the work by \citet{sheth08} was published. Since then, a possible solution to this discrepancy was presented by \citet{cameron10} and later by \citet{melvin14}. They found that this trend is more acute for low-mass galaxies ($\log(M_{\star}/M_{\sun}) < 10.34$) than for high-mass galaxies ($\log(M_{\star}/M_{\sun}) \ge$ 10.64). This is consistent with an scenario where more massive, dynamically cold, stellar discs are already settled during the first Gyrs of the history of the Universe, and therefore they have had enough time to develop bars and reach the observed low-redshift bar fraction early in time. The recent discovery of six massive barred spirals at $z > 1$ by \citet{guo23} also points in this direction, setting the formation of cold disc very early in the history of the Universe.

Three common caveats are always associated with the identification of bars at high redshift: i) limited spatial resolution and evolution of the physical scale with redshift; ii) the morphological K-correction; and iii)  surface brightness dimming. The impact of spatial resolution in the detection of bars is a widely discussed topic even when comparing galaxy samples in the local Universe. The common agreement is that bars can be detected if they are at least about two times the full width at half maximum  $(\sim$2 $\times$  FWHM) of the  point spread function (PSF) in terms of the radius \citep{aguerri09}. \citet{erwin18} showed that detecting the short end of the bar length distribution might be a critical point to compare bar fractions from different studies, but the vast majority of HST studies have explored bars in the rest-frame optical light. This might also have implications when comparing bar fractions since bars are stellar structures (easily detected at redder wavelengths) and dust effects can obscure small bars. Finally, a generally untouched problem is the cosmological surface brightness dimming. This might affect the observability of the outer disc making it more difficult to separate stellar bars from elliptical galaxies \citep{melvin14}. Nevertheless, the effects of the surface brightness dimming on the detection of bar-built structures such as boxy/peanut bulges was studied by \citet{kruk19}, without finding significant discrepancies between the local Universe and $z \sim$0.4.

This paper represents the first attempt to use the new capabilities of JWST to measure the fraction of barred galaxies in a cluster at $z$=0.39. The characteristics of the NIRCam imaging of the Early Release Observations (ERO) of the SMACS0723 cluster overcome previous issues related to spatial resolution, bar identification at rest-frame wavelengths, and depth of the observations. Therefore, we are able to provide a robust estimation of the bar fraction and pave the way for future studies of bar evolution.

This paper is organized as follows. Section~\ref{sec:smacs} describes our fiducial sample of galaxy cluster members. Section~\ref{sec:visual} shows the process of visually classifying whether (or not) cluster member galaxies host a bar. Section~\ref{sec:fraction} highlights the main results of our study. Section~\ref{sec:discussion} places our results in the context of bar fraction evolution with redshift and environment. Finally, Section~\ref{sec:conclusions} provides a summary of our main conclusions. Throughout the paper, we assume a flat cosmology with $\Omega_m$ = 0.3, $\Omega_{\lambda}$ = 0.7, and a Hubble constant of $H_0$ = 70 km s$^{-1}$ Mpc$^{-1}$.

\section{SMACS0723 and the cluster membership}
\label{sec:smacs}

The galaxy cluster SMACSJ0723.3-7323 (hereafter, SMACS0723) is part of the southern extension of the MACS sample \citep{ebeling10,reppebeling18}. \citet{mahler22}, using the ROSTAT package \citep{beers90} derived a cluster redshift of $z$=0.3877 using a sample of 26 spectroscopically confirmed members of the cluster. They also derived a cluster velocity dispersion of $\sigma \sim $1180$\pm$160 km s$^{-1}$. The cluster total mass estimated by Planck is 8.39$\times$10$^{14}$ M$_{\sun}$ \citep{coe19}. Using this value and the equations given by \citet{coe10} we derived the cluster viral radius (conventionally defined as the radius within which the mean density is 200 times the background density) as $r_{200}$ = 1.95 Mpc = 6.15 arcmin.

The SMACS0723 cluster has been observed as part of the RELICS program \citep{coe19}. They observed a sample of 41 massive galaxy clusters with HST (PI: D. Coe), and the Spitzer Space Telescopes (PI: M. Bradac). Deep observations (26.5 AB mag) were obtained for these clusters in 7 HST bands: F435W, F606W, F814W with the Advanced Camera for Surveys (ACS), and F105W, F125W, F140W, and F160W with the Wide Field Camera Three (WFC3). The RELICS data products for SMACS0723 include reduced and color images, photometric catalogs generated with SExtractor \citep{bertinarnouts96}, and photometric redshifts computed with the Bayesian Photometric Redshifts code \citep[BPZ]{benitez00}. These are publicly available through the RELICS website\footnote{https://relics.stsci.edu/data.html}.

\begin{figure}
   \centering
   \includegraphics[width=0.49\textwidth]{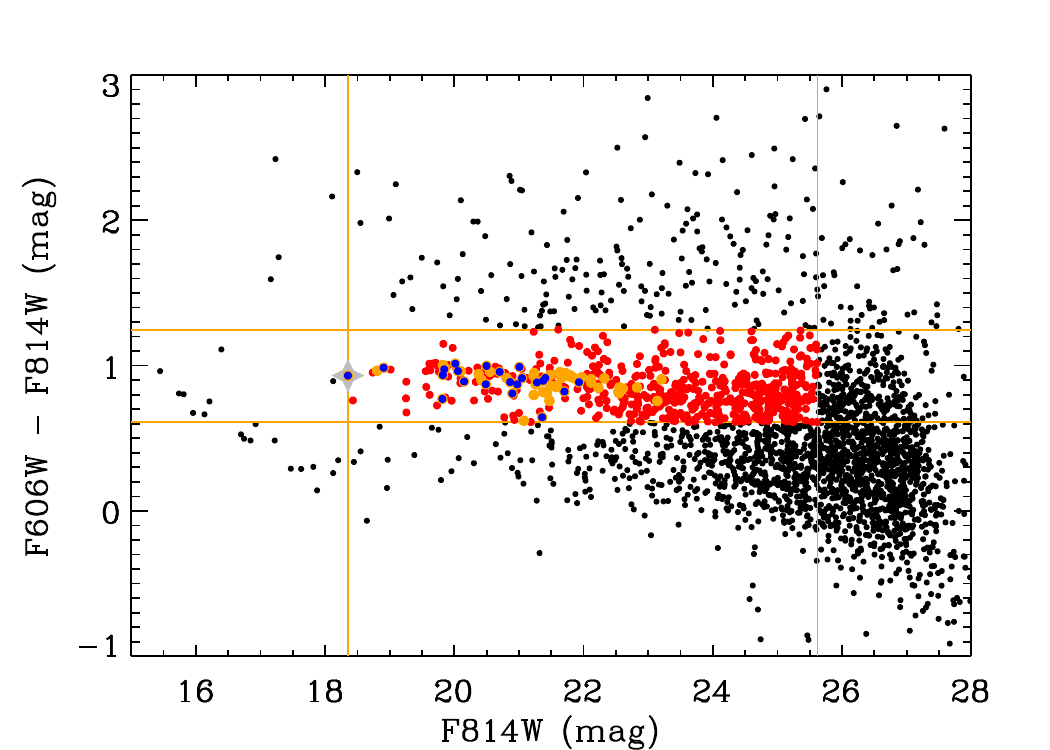}
      \caption{Color (F606W-F814W) vs. magnitude (F814W) diagram of all sources present in the RELICS catalogue (Black points). Possible cluster members selected using both the red sequence colors (0.6134 $<$F606W-F814W$<$1.2478 mag; orange horizontal lines) and the magnitude limits (25.62$<m_{\rm{F814W}}<$18.35 mag; orange vertical lines) are shown in red. The subsample of 22 spectroscopically confirmed cluster members by \citet{mahler22} are shown in blue. The subsample of 61 spectroscopically confirmed cluster members by \citet{noirot22} are shown in orange. The brightest cluster galaxy (BCG) of the cluster is represented with a grey star.
              }
         \label{fig:colormag}
   \end{figure}

In order to select the galaxy cluster members we first used a criterion based on the color-magnitude relation. Galaxy clusters are known to display a well-defined red-sequence that can be used to photometrically identify cluster members \citep{gladders05}. After inspecting all possible combinations of colors available in the RELICS catalogue, we found that the red-sequence was better defined when using the F606W and F814W filters. Using a similar analysis, \citet{golubchik22} identified a sample of 130 cluster members with magnitudes brighter than 23 in the F814W band. Using the same magnitude criteria we found 116 cluster members, which we consider a good agreement since we might have applied a different color cut. Then, we used the E-MILES library of single stellar population models \citep{vazdekis16} to derive the colors of an old (14 Gyr), metal-rich ([M/H]=0.4), and high [$\alpha$/Fe]=0.4 model (red galaxy) and a young (1 Gyr), solar-metallicity ([M/H]=-0.25), solar [$\alpha$/Fe]=0 abundance (blue galaxy) at the redshift of the cluster ($z$ = 0.39). Figure~\ref{fig:colormag} shows the color-magnitude relation for all galaxies in the ACS/RELICS field of view (FoV), with the two colors defining the red sequence of SMACS0723 (0.6134 $<$F606W$-$F814W$<$1.2478). Both reddest and bluest galaxies can also be hosted by the cluster due to dust reddening (former) or recent star formation (latter), but both effects are not very common in massive clusters such as SMACS0723. This preliminary selection provided us with an initial sample of 590 galaxies with absolute magnitudes F814W < $-16$ mag ($m_{\rm{F814W}}$ = 25.6). This low magnitude limit was set to avoid cluster membership confusion at the dwarf end of the luminosity function and because in previous works we find no bars at fainter magnitudes \citep{mendezabreu12}. We also imposed a limit at the bright end of the color-magnitude relation. This was set to $m_{\rm{F814W}}$ = 18.35~mag which corresponds to the magnitude of the brightest cluster galaxy.

In order to check the reliability of our red sequence selection process, we show in Fig.~\ref{fig:colormag} the position of the 22 spectroscopically confirmed galaxies as cluster members in \citet[][Table 1]{mahler22}. We also included in Fig.~\ref{fig:colormag} a sample of 61 spectroscopically selected member galaxies from the recent database of \citet[][]{noirot22}. These were chosen by imposing a simple redshift cut $0.36<z<0.42$. We found that all SMACS0723 spectroscopic members are selected as possible cluster members following our color selection.

In this work, we use the Early Release Observations (ERO) of the SMACS0723 cluster \citep[proposal ID: 27361; ][]{pontoppidan22}. Observations were taken on 7 June 2022,  using nine dither positions to optimize image quality, exposures of a total of 7\,537~s per filter to achieve a point source sensitivity of AB $\sim$ 29.8 mag ($\sim$ 3 magnitudes deeper than RELICS), and the MEDIUM8 readout pattern to minimize detector read noise. The public release includes calibrated mosaics in six broad-band NIRCam filters (i.e., F090W, F150W, F200W, F277W, F356W, and F444W), available on the Mikulski Archive for Space Telescopes (MAST). Since the analysis of the visual morphology of galaxies (bar detection) does not require absolute flux calibration or high-precision astrometry, a careful visual inspection of the public dataset reveals a good quality of the automatic reduction. In particular, we created postage stamps of each galaxy member of the SMACS0723 cluster using the F200W filter, since it provides the best spatial resolution (0.031 arcsec/px; PSF FWHM of 0.066 arcsec) and sensitivity. We notice here that all photometric information about the galaxies used in this paper was derived from HST data mainly because a more robust photometric calibration and catalogue selection was available when this work was in progress.

The JWST/NIRCam FoV observes a 9.7 arcmin$^2$ field with a $\sim$44 arcsec gap separating two 2.2 arcmin $\times$ 2.2 arcmin areas. NIRCam observations of the SMACS0723 cluster were taken with one camera centered on the cluster, and another on an adjacent field. Therefore, they cover a smaller area of the cluster with respect to the RELICS HST/ACS imaging (3.36 arcmin $\times$ 3.36 arcmin). We found that 300 galaxies out of the initial 590 were present in the NIRCam imaging of SMACS0723. This final number already includes the removal of some obvious stars and duplicated object in the initial RELICS sample. The final number of cluster members analysed in this study also includes the following cuts: i) in order to avoid non-resolved point sources we imposed a stellarity parameter lower than 0.9 (see RELICS catalogue for details) and the condition that they are not visually classified as point source (see Sect.~\ref{sec:visual}); ii) galaxies should be relatively face-on ($\epsilon < 0.5$) to avoid projection problems and to be comparable with previous studies \citep[e.g.,][]{mendezabreu12}; and iii) galaxies should have a photometric redshift (see RELICS catalogue for details) $z < $ 1 to avoid contamination from background galaxies. We checked that all spectroscopically confirmed members satisfy this condition. Our final sample of SMACS0723 cluster members consist of 188 galaxies. 

\begin{figure}
   \centering
   \includegraphics[width=0.49\textwidth]{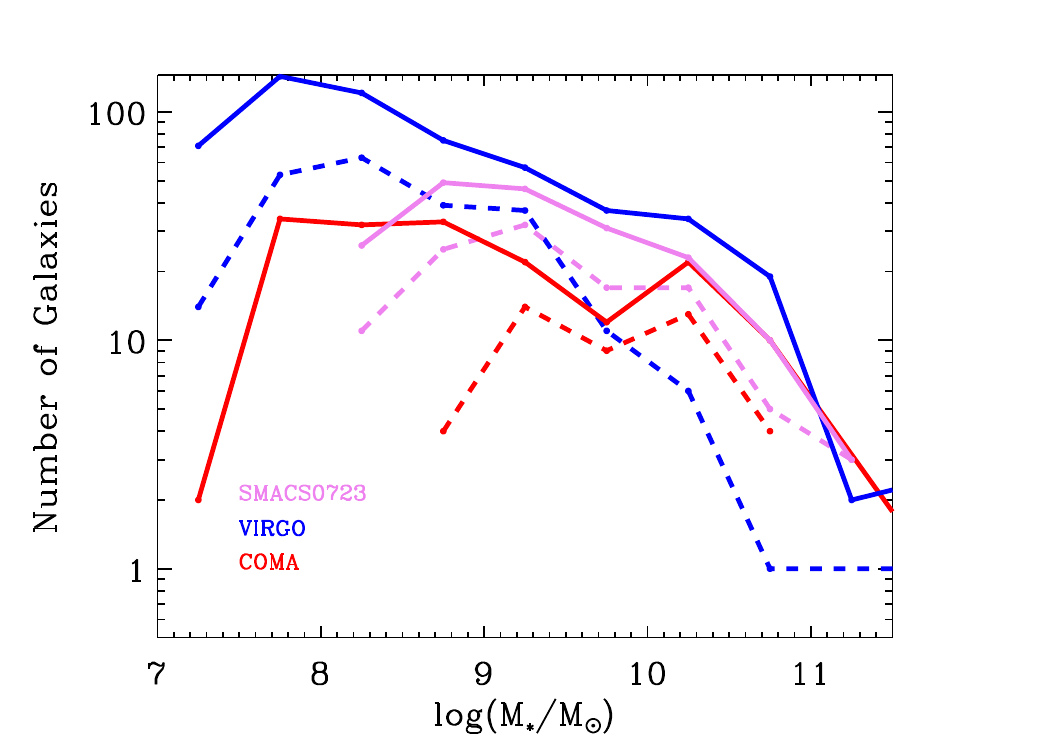}
      \caption{Total number of galaxies (solid lines) and disc galaxies (dashed lines) as a function of the stellar mass for the Coma (red), Virgo (blue), and SMACS0723 (violet) clusters. The values for both Coma and Virgo clusters were obtained from \citet{mendezabreu12} and they are described in Sect.~\ref{sec:fraction}.}
         \label{fig:mass}
   \end{figure}

\begin{figure*}
   \centering
   \includegraphics[width=0.8\textwidth]{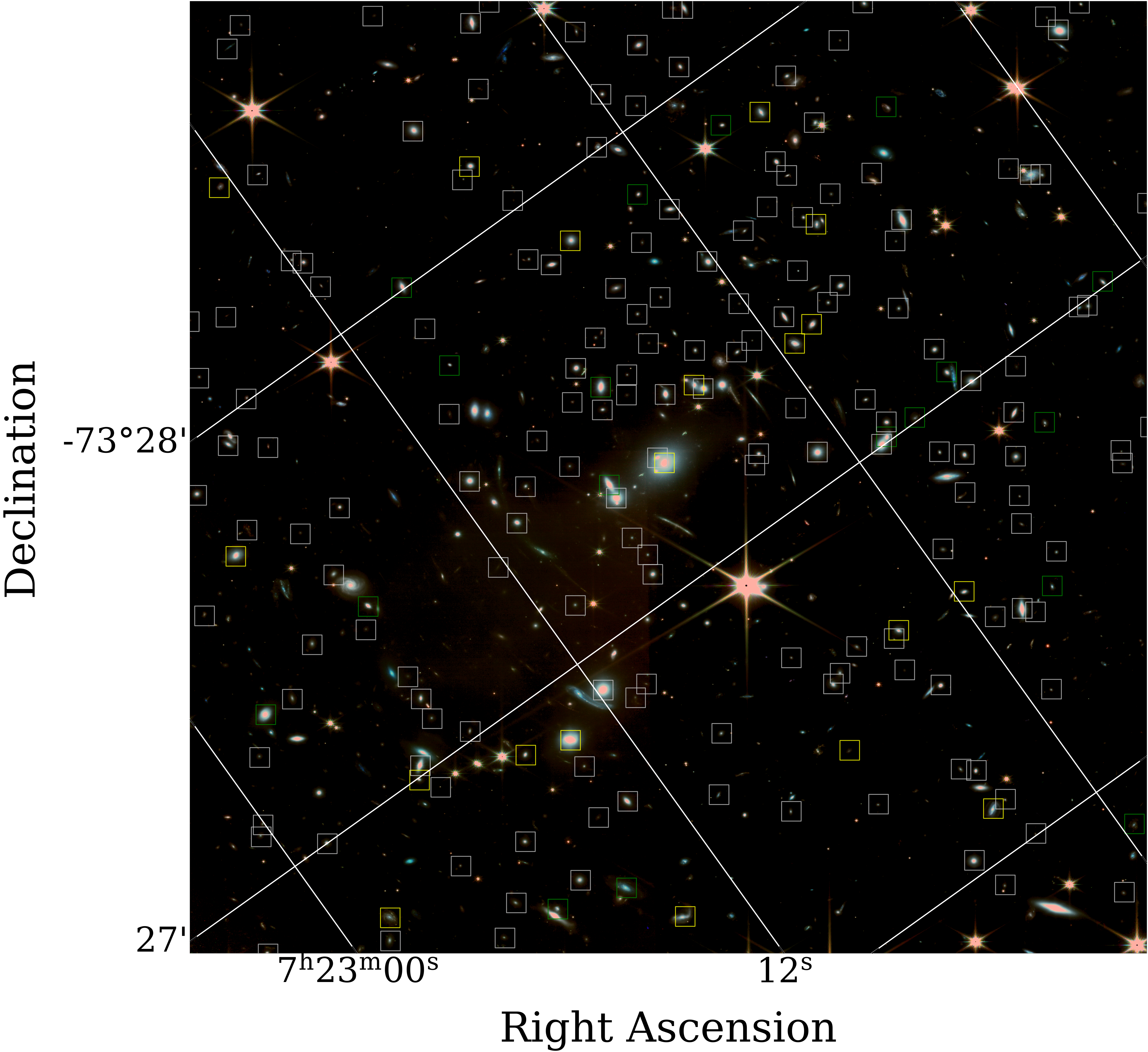}
      \caption{JWST color image of SMACS0723 showing the spatial distribution of cluster members confirmed spectroscopically (green) and using our color-magnitude cuts (white). This image was produced from our reduced data products via a composite of data in 3 bands: F090W, F150W, F200W. F090W was assigned blue colours, F150W green, and F200W red.
              }
         \label{fig:cluster}
   \end{figure*}

Figure~\ref{fig:mass} shows the stellar mass distribution (as computed in Sect.~\ref{sec:fraction}) of both all SMACS0723 cluster members and only those galaxies classified as discs (see Sect. \ref{sec:visual}). The stellar mass distribution for the nearby Coma and Virgo clusters, obtained by \citet{mendezabreu12}, are also represented for comparison. The spatial distribution of the SMACS0723 cluster members is shown in Fig.~\ref{fig:cluster}.

\section{Galaxy morphological classification}
\label{sec:visual}

   \begin{figure*}
   \centering
   \includegraphics[width=1\textwidth]{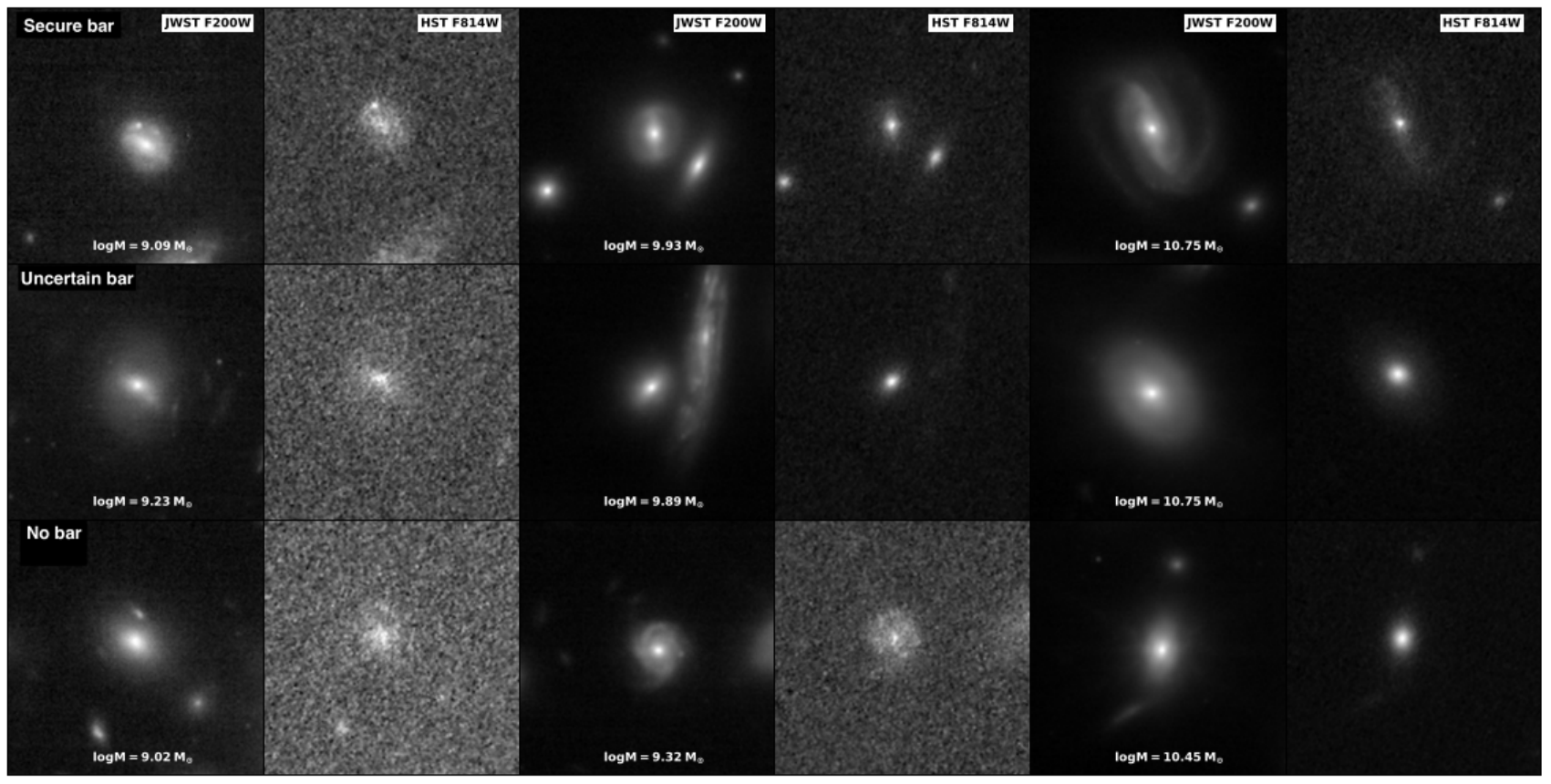}
      \caption{Examples of JWST SMACS0723 galaxies in the F200W filter, classified in this study: secure bars (first row), uncertain bars (second row), unbarred disc galaxies (third row) compared to the HST F814W images. The galaxies are ordered from left to right by increasing stellar mass. The postage stamps are 4.65 $\times$ 4.65 arcsec in size.}
         \label{fig:examples}
   \end{figure*}

To identify bars in the JWST cutout images of the galaxies more efficiently, we set up a private project using the Zooniverse Panoptes Project Builder. 

We created \textsc{fits} cutouts of the cluster members and converted them to 424 $\times$ 424 pixels \textsc{jpeg} postage stamps, applying an arcsinh stretch to the images. Following  \citet{costantin23}, we also derived parametric and non-parametric morphology of the sample galaxies using \textsc{statmorph} \citep{rodriguezgomez19}. We focused on the F200W image (CAS parameters, 1-component S\'ersic model and residuals) and show this output to the classifiers. All of these images were informative on whether the galaxy hosts a bar or not.

We set up a simple workflow in the Zooniverse Panoptes framework for classification. First, we filtered on whether the galaxy has been classified before (to remove potential duplicates due to shredding or multiple identifications of the same object in the photometric catalogue). Secondly, we classified the orientation of the objects, in order to remove edge-on cases, where the bar identification is difficult. Then we classified the global morphology of the galaxies into four classes: spheroid, disc, irregular or point source. In cases where the galaxy was identified to be a disc or irregular type, we classified whether a bar was present in the galaxy image. In total, 188 galaxies were classified by all three authors for the presence of a bar, with a total of 564 classifications.

We then aggregated the classifications. If all three classifiers agreed that there was a bar present in the galaxy, we classified the galaxy as having a secure bar. If at least one of the three classifiers identified the galaxy as being barred, the galaxy was classified as hosting an uncertain bar. Examples of galaxies with secure, uncertain, and unbarred discs classified in this work are shown in Figure \ref{fig:examples}, in comparison with the HST ACS F814W images. In total, there are  20 secure bars and 15 uncertain bars out of 188 galaxies in the sample. To account for both secure and uncertain bars in the sample, in the following analysis the lower error bars include the secure bars and binomial errors, while the higher error bars include the secure and uncertain bars, as well as binomial errors. 

\section{The bar fraction in the SMACS0723 cluster}
\label{sec:fraction}

Our analysis of the bar fraction in the SMACS0723 galaxy cluster includes two different definitions: we derived the ordinary bar fraction, f$_{D}$, (as it is usually calculated using only disc galaxies) and the overall bar fraction, f$_{T}$, (calculated using all galaxies independently of their Hubble type). Since bars can only be triggered in discs, f$_{D}$, has been historically deemed as the correct way of computing the bar fraction. However, the visual morphological separation between massive non-barred lenticulars and elliptical galaxies is very difficult and introduces a large uncertainty in f$_{D}$. On the other hand, f$_{T}$ avoids this problem and allows us to probe a larger range of luminosities and masses than f$_{D}$, but it assumes that the luminosity/mass distribution of elliptical versus disc galaxies is the same in the different samples under comparison, which might not be the case when comparing clusters with very different masses or when comparing different galactocentric regions of the clusters. We used both in our analysis to provide a more complete picture of the bar fraction and to compare them with local studies using similar quantities.

\begin{figure*}
   \centering
   \includegraphics[bb= 0 0 470 350,width=0.49\textwidth]{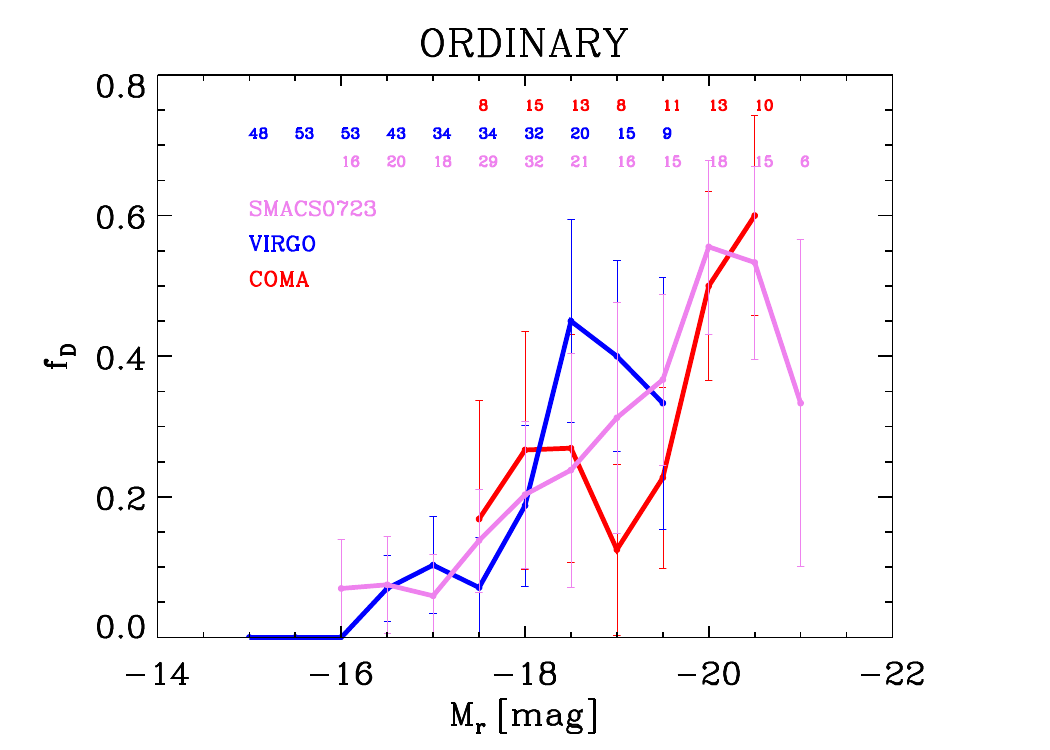}
    \includegraphics[bb= 0 0 470 350,width=0.49\textwidth]{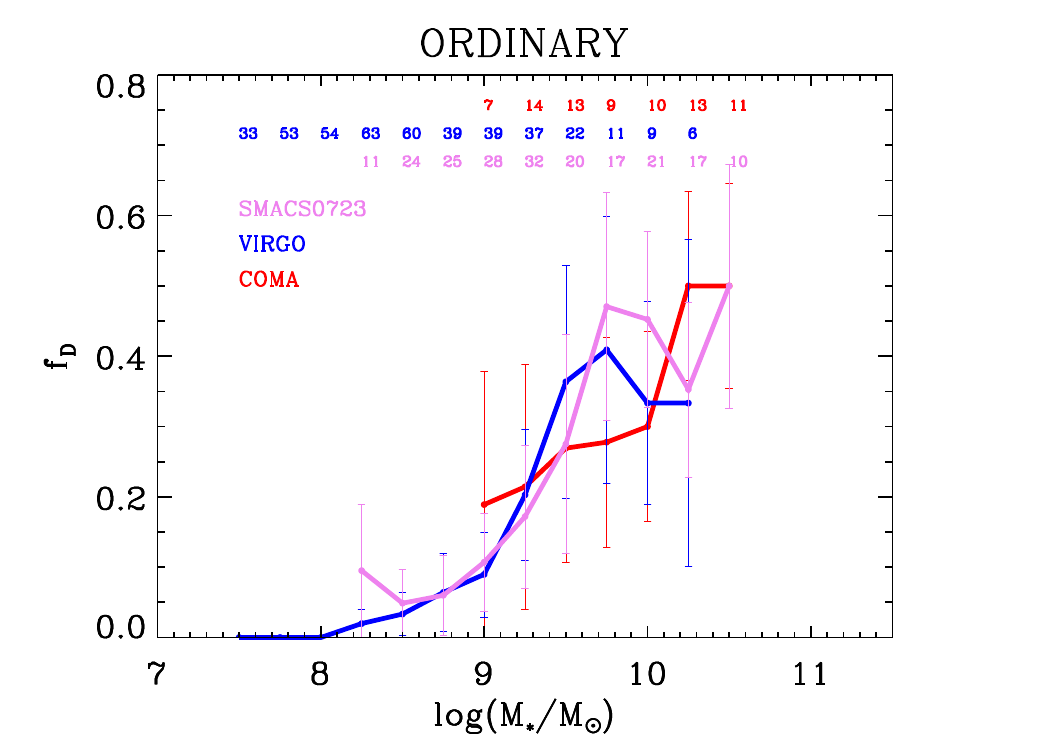}
    \includegraphics[bb= 0 0 470 350,width=0.49\textwidth]{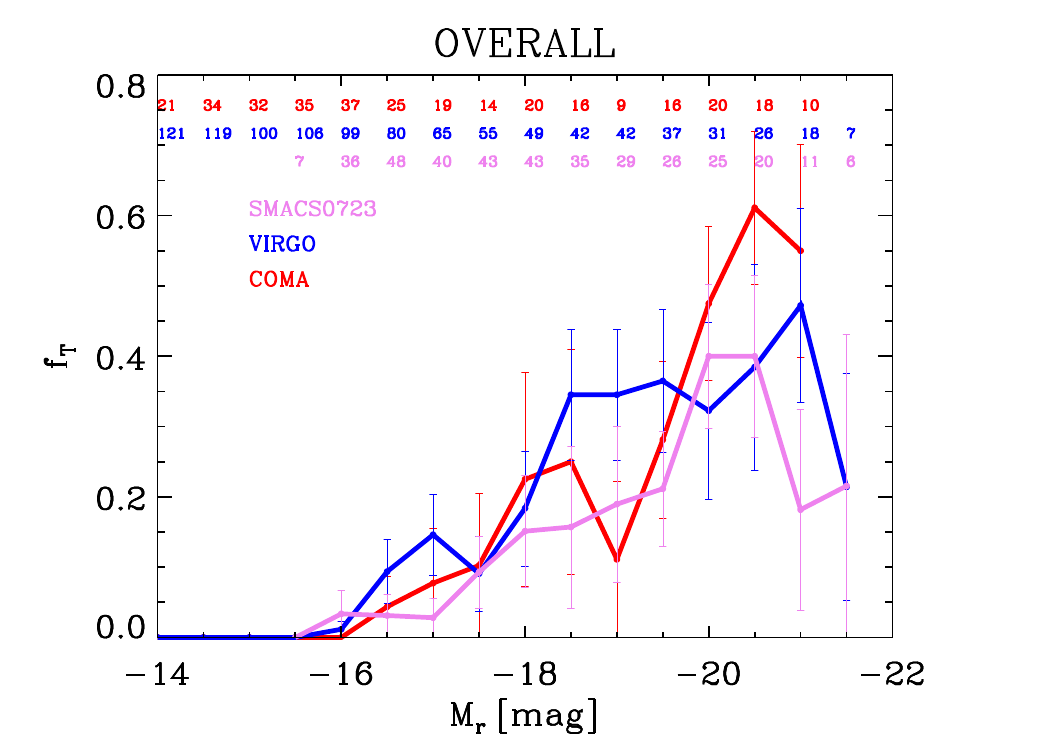}
    \includegraphics[bb= 0 0 470 350,width=0.49\textwidth]{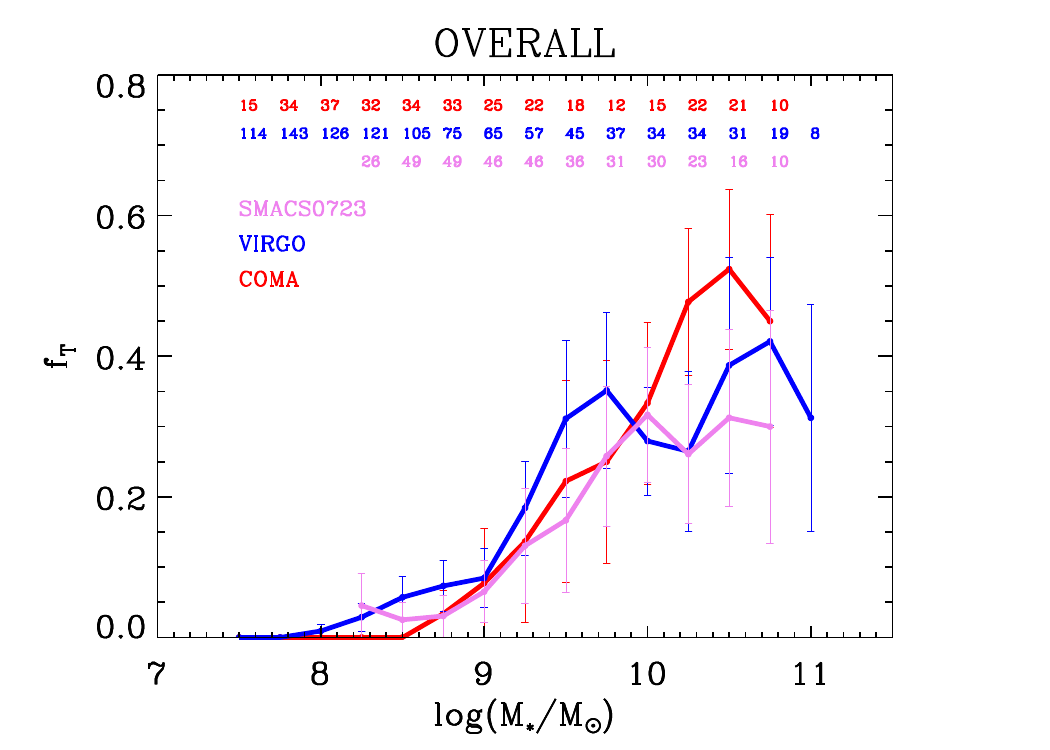}
      \caption{Bar fraction for the Coma (red), Virgo (blue), and SMACS0723 (violet) clusters as a function of both the absolute $r-$band magnitude (left panels) and the stellar mass (right panels). The bar fractions have been computed using both only disc galaxies (ORDINARY, upper panels) and all galaxies (OVERALL, lower panels). The number of galaxies in each bin is shown at the top of each panel coloured accordingly. As explained in Sect.\ref{sec:fraction}, bins are represented either every 0.5 mag or 0.25 dex, but the fraction (and therefore the number of galaxies) is averaged over bins of 1 mag and 0.5 dex for the magnitudes and masses, respectively.
              }
         \label{fig:barfrac}
   \end{figure*}

Figure~\ref{fig:barfrac} shows f$_{D}$ and f$_{T}$ as functions of both the SDSS $r$-band absolute magnitude and stellar mass of the galaxies in the SMACS0723 cluster. We transformed the magnitudes obtained from the RELICS catalogue using the HST/ACS F814W filter to the SDSS $r$-band system by using the E-MILES library of single stellar population (SSP) models \citep{vazdekis16}. To this aim, we first assumed our galaxies to be represented with four extreme SSP properties (as described in Sect.~\ref{sec:smacs}: i) an old (14 Gyrs) and metal-rich ([M/H]=0.4); ii) an old (14 Gyrs) and metal-poor ([M/H]=-0.25); iii) a young (1 Gyr) and metal-rich ([M/H]=0.4); and iv) a young (1 Gyr) and metal-poor ([M/H]=-0.25). All these representing extreme cases of the possible galaxy population in our cluster. Then, we computed the magnitudes of these SSP models for both the F814W and $r-$band filters at the redshift of the cluster $z$=0.39 and $z$=0, respectively. The differences obtained between the magnitudes at different bands (redshift) will provide us with the typical correction for each particular SSP. We finally computed the mean difference of the 4 SSP models to be  0.155 mag. This correction was then applied to transform F814W magnitudes into $r$-band ones. The same procedure was carried out to transform the F606W filter into the SDSS $g-$band obtaining a correction factor of 0.058 mag. This was necessary to compute the galaxy stellar masses using the prescriptions given by \citet{zibetti09}. Figure~\ref{fig:barfrac} also shows the bar fractions for the Virgo and Coma cluster as derived in \citet{mendezabreu12}. The three clusters are now directly comparable since the magnitudes, colors, and stellar masses have been computed in the same way. To avoid issues related to the bin size, we applied a moving-average (boxcar) smoothing over the histograms using box widths of both 1 mag and 0.5 dex and steps of 0.5 mag and 0.25 dex in magnitude and mass, respectively. The number of galaxies in each bin is shown at the top of each panel in Fig.~\ref{fig:barfrac}. It is worth noting that due to our smoothing method some galaxies can be counted in two adjacent bins. The bar fraction errors are calculated by considering only the secure bars and both the secure and uncertain bars, respectively, and including their statistical uncertainties. The latter were computed by estimating the confidence intervals on binomial population proportions following the prescriptions by \citet{cameron11}.

Figure~\ref{fig:barfrac} shows that, independently on how we compute it, the bar fraction for the three clusters is a strong function of galaxy luminosity (stellar mass) as already discussed by several authors \citep{mendezabreu10a,nairabraham10,mendezabreu12,erwin18,kruk18}. The overall bar fraction (f$_{T}$) in all clusters shows a peak around M$_{r}$ $\sim -20.5$  in absolute magnitude and $\log(M_{\star}/M_{\sun})$ $\sim$10.5,  followed by a similar decrease towards fainter (low-mass) galaxies. The observed trends are similar when using the ordinary bar fraction (f$_{D}$), but at higher luminosities and masses, the Virgo cluster presents a lack of discs in our sample that makes uncertain the computation of the bar fraction. We calculated the weighted mean, peak value, and corresponding errors in magnitude (and mass) of the bar fraction distributions of the three cluster by performing a series of 1000 Monte Carlo simulations taking into account the confidence intervals. These results are shown in Table~\ref{table:barfractable}.

\begin{table*}[!t]  
\label{table:barfractable}
\begin{center}
\caption{Weighted mean and peak value of the luminosity and mass bar fraction distributions \label{tab:statistics}}  
\begin{tabular}{cclrrr}    
\hline  
\hline   
\multicolumn{1}{c}{Bar Fraction Distribution} & 
\multicolumn{1}{c}{Statistical Parameter} &  
\multicolumn{1}{c}{Galaxy Property} & 
\multicolumn{1}{c}{SMACS0723} &   
\multicolumn{1}{c}{Virgo} & 
\multicolumn{1}{c}{Coma} \\   
\multicolumn{1}{c}{(1)} & \multicolumn{1}{c}{(2)} &   
\multicolumn{1}{c}{(3)} & \multicolumn{1}{c}{(4)} &   
\multicolumn{1}{c}{(5)} & \multicolumn{1}{c}{(6)} \\   
\hline\rule{0pt}{2ex}
                           &                        & Luminosity &     $-19.4$$\pm$0.3  &     $-18.5$$\pm$0.2  &     $-19.4$$\pm$0.2 \\[-1ex]
                           & \raisebox{1.5ex}{Mean} & Mass       &      9.8$\pm$0.1   &       9.7$\pm$0.1  &       9.9$\pm$0.1 \\
                     
\raisebox{1.5ex}{Ordinary}  &                       & Luminosity &     $-20.2$$\pm$0.6  &     $-18.9$$\pm$0.4  &     $-20.2$$\pm$0.6 \\[-1ex]
                           & \raisebox{1.5ex}{Peak} & Mass       &      10.1$\pm$0.4  &       9.9$\pm$0.3  &       10.2$\pm$0.4 \\[1ex]
\hline\rule{0pt}{2ex}
                           &                        & Luminosity &     $-19.7$$\pm$0.3 &     $-19.5$$\pm$0.2 &     $-19.7$$\pm$0.2 \\[-1ex]
                           & \raisebox{1.5ex}{Mean} & Mass       &      10.0$\pm$0.1 &       9.9$\pm$0.1 &      10.1$\pm$0.1 \\
\raisebox{1.5ex}{Overall} &                         & Luminosity &     $-20.4$$\pm$0.7 &     $-20.4$$\pm$0.8 &     $-20.6$$\pm$0.4 \\[-1ex]
                           & \raisebox{1.5ex}{Peak} & Mass       &      10.3$\pm$0.4 &      10.4$\pm$0.4 &      10.5$\pm$0.2 \\
\hline    
\end{tabular}
%
\tablefoot{Luminosities and masses are given in SDSS $r$-band magnitudes and $\log{(M_{\star}/ M_{\sun})}$, respectively.
}
\end{center}
\end{table*}

In the low luminosity and mass range (M$_r > -18.5$~mag; $\log(M_{\star}/M_{\sun}) \leq$ 9.5) the bar fraction distribution in the three clusters is essentially the same. At M$_r \sim -18.5$~mag ($\log(M_{\star}/M_{\sun}) \sim$ 9.55), the typical bar fraction in the three clusters is $\sim$ 30\% dropping to 0\% at M$_r \sim$ $-16$ mag. The mean bar fractions in this luminosity/mass range are shown in Table~\ref{tab:bars}. At intermediate luminosities and masses ($-18.5 \geq$ M$_r \geq -20$~mag;  9.5$\leq \log(M_{\star}/M_{\sun}) \leq$ 10.25), the bar fraction in the Virgo cluster shows a secondary peak in all distributions (f$_{T}$ and f$_{D}$) which is not clear in either SMACS0723 nor Coma. Actually, the Coma cluster shows a dip in the bar fraction when looking at the magnitude distributions of both f$_{T}$ and f$_{D}$, however, it does not appear in the mass distribution so we believe it might be a statistical fluctuation due to low number statistics. At high luminosities and masses (M$_r < -20$~mag;  $\log(M_{\star}/M_{\sun}) >$ 10.25), the discs based bar fractions of SMACS0723 is lower than in Coma. When considering all galaxies, the differences are even more acute with the bar fraction of SMACS0723 being lower than Virgo, and Coma showing the highest values. The mean bar fractions in this luminosity/mass range are shown in Table~\ref{tab:bars}. The differences observed at high luminosities and masses between f$_{D}$ and f$_{T}$ might indicate a different fraction of ellipticals versus disc galaxies in the three clusters, rather than an actual difference in the bar fraction. In this case, the ellipticals-to-discs fraction should be larger in SMACS0723 and Virgo than in Coma.

\begin{table}
\caption{Mean bar fraction in the three luminosity and mass intervals}             
\label{tab:bars}      
\centering                          
\begin{tabular}{l c c c}        
\hline\hline                 
Range                                  & Coma & Virgo  & SMACS0723 \\    
\hline                        
                  & OVERALL & &\\ 
Low luminosity                          & 11\%   & 11\% & 7\% \\  
\vspace{0.2cm}
Low mass                                & 8\%   & 7\% & 6\% \\

Interm. luminosity                     & 28\%   & 34\% & 24\% \\  
\vspace{0.2cm}
Interm. mass                           & 32\%   & 31\% & 25\% \\

High luminosity                        & 58\%   & 36\% & 27\% \\
High mass                              & 49\%   & 37\% & 31\% \\
\hline 
                  & ORDINARY & &\\
Low luminosity                          & --   & 11\% & 11\% \\
\vspace{0.2cm}
Low mass                                & --   & 8\% & 10\% \\

Interm. luminosity                     & 28\%   & 39\% & 37\% \\  
\vspace{0.2cm}
Interm. mass                           & 34\%   & 36\% & 39\% \\

High luminosity                         & 60\% & --   & 43\% \\
High mass                              & 50\% & --   & 50\% \\
\hline                                   
\end{tabular}
\tablefoot{High-, intermediate-, and low-luminosity(mass) ranges correspond to:  M$_r < $ $-20$ ($\log(M_{\star}/M_{\sun}) >$ 10.25), $-18.5$ $\geq$ M$_r$ $\geq$ $-20$ (9.5 $\leq$ $\log(M_{\star}/M_{\sun}) \leq$ 10.25), and M$_r >$ $-18.5$ ($\log(M_{\star}/M_{\sun}) <$ 9.5), respectively.}
\end{table}

\begin{figure*}
   \centering
   \includegraphics[bb= 0 0 470 350,width=0.49\textwidth]{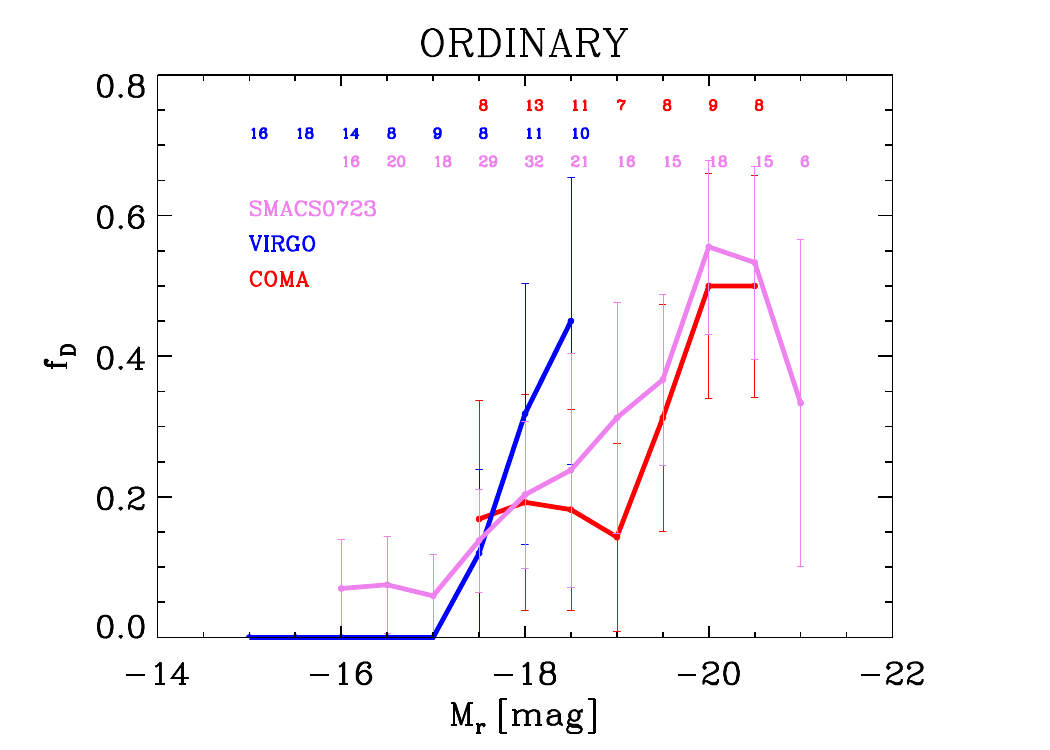}
    \includegraphics[bb= 0 0 470 350,width=0.49\textwidth]{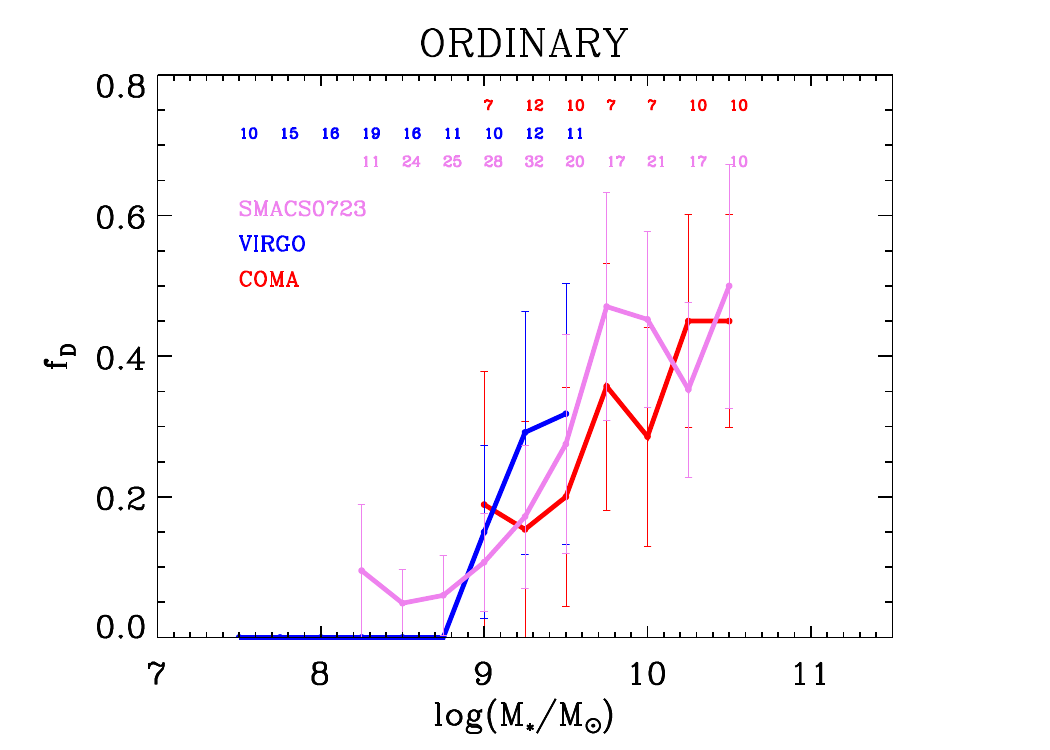}
      \caption{Bar fraction for the Coma (red), Virgo (blue), and SMACS0723 (violet) clusters as a function of both the absolute $r-$band magnitude (left panel) and the stellar mass (right panel). The bar fractions have been computed using only disc galaxies (ORDINARY) within a clustercentric radius of 0.3$\times$r$_{200}$ (see text for details). The number of galaxies in each bin is shown at the top of each panel coloured accordingly. As explained in the text, bins are represented either every 0.5 mag or 0.25 dex, but the fraction (and therefore the number of galaxies) is averaged over bins of 1 mag and 0.5 dex for the magnitudes and masses, respectively.
              }
         \label{fig:barfracinner}
   \end{figure*}

One possible caveat when comparing the results of the SMACS0723 cluster with those of Virgo and Coma is the different spatial coverage. Our results on the SMACS0723 cluster have been obtained using the NIRCam photometry centred in the brightest cluster galaxy. Therefore, we are mapping a clustercentric radius of $\sim$ 0.3$\times$ r$_{200}$. Considering values of r$_{200}$=2.86 Mpc \citep{lokasmamon03} and r$_{200}$=2.86 Mpc \citep{ferrarese12} for Coma and Virgo, respectively, we limited their samples to match the r $\sim$ 0.3$\times$r$_{200}$ spatial coverage in all three clusters. The ordinary bar fractions on these restricted samples are shown in Fig.~\ref{fig:barfracinner}. The results discussed previously do not change when considering the same spatial coverage despite the more limited ranges they probe.

A potential bias that might also affect the comparison between the bar fractions of the nearby clusters (Virgo and Coma) and SMACS0723 is the different wavelength range used to identified the bars. In order to take advantage of the full capabilities of NIRCam, we used the F200W images that best represent the combination of both image quality and depth of the JWST observations. The F200W band at a $z$=0.39 correspond to an intermediate wavelength between the J-band and  H-band at rest-frame; therefore, we are mapping the near-infrared population of bars in SMACS0723. However, the study of the Virgo and Coma clusters were carried out using optical imaging in the SDSS r-band and ACS-HST F814W images, respectively. In order to quantify the impact of the different wavelength range in the identification of bars, we performed a further visual inspection of the barred galaxies detected in the F200W images, but this time using the F090W NIRCam filter. The F090W band at a $z$=0.39 correspond to approximately to the SDSS r-band at rest-frame. We found 10 secure bars and 18 uncertain bars in the sample corresponding to a global bar fraction of 25\%, instead of the 31\% derived using the F200W. The lower bar fraction in the rest-frame optical with respect to the rest-frame infrarred is expected and it has been also reported in low-redshift studies \citep[e.g.,][]{erwin18}. Figure \ref{fig:barfraccomp} displays the comparison of the bar fraction derived with both NIRCam filters (F200W and F090W) as a function of the stellar mass. The trend of the bar fraction with stellar mass is the same as discussed before, but the bar fraction is smaller at all masses. In the low-luminosity and mass range (M$_r > -18.5$~mag; $\log(M_{\star}/M_{\sun}) \leq$ 9.5), the bar fractions between the two filters is similar and therefore our results do not change. At intermediate luminosities and masses ($-18.5 \geq$ M$_r \geq -20$~mag;  9.5$\leq \log(M_{\star}/M_{\sun}) \leq$ 10.25), we found the largest differences depending on the filter, even if the overall shape of the distribution is the same. At high luminosities and masses (M$_r < -20$~mag;  $\log(M_{\star}/M_{\sun}) >$ 10.25), the bar fraction is lower in the F090W, this will reinforce our previous result that the bar fraction for high-mass galaxies in SMACS0723 is lower than in the Coma cluster.

   \begin{figure}
   \centering
   \includegraphics[width=0.49\textwidth]{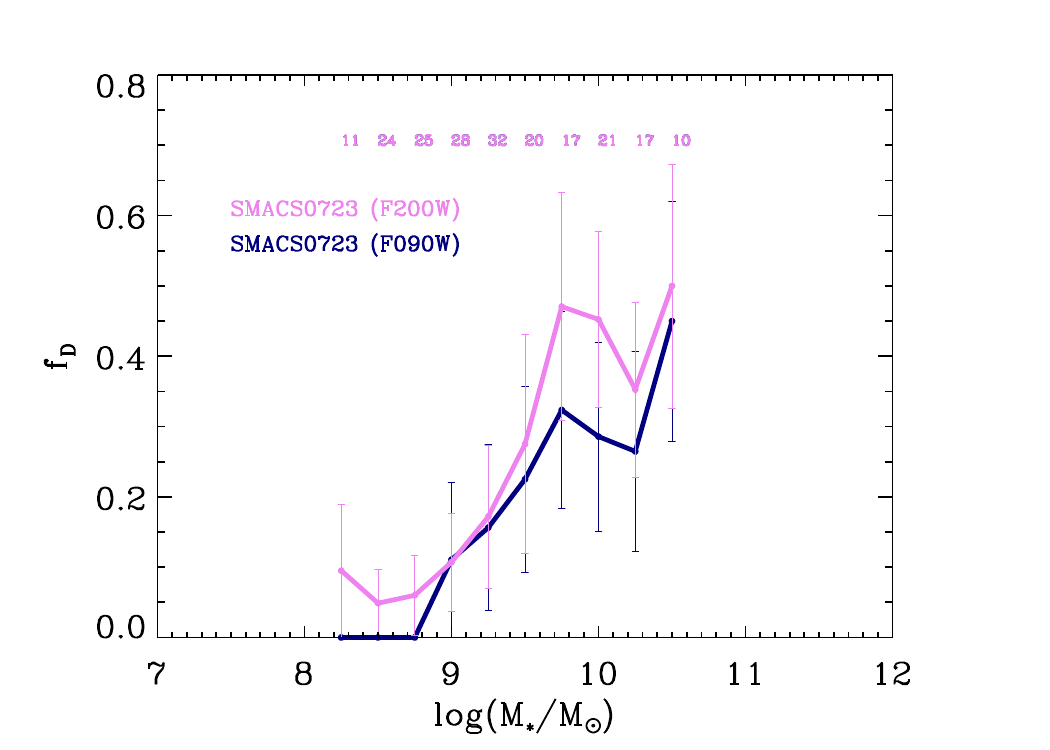}
      \caption{Bar fraction for the SMACS0723 cluster as a function of stellar mass computed using the F090W band (dark blue) and the F200W band (violet) images to identify bars. The bar fractions have been computed using only disc galaxies (ORDINARY). The labels and binning scheme is the same as in Fig. \ref{fig:barfrac}.
              }
         \label{fig:barfracomp}
   \end{figure}
  
Another critical aspect when comparing the bar fractions of different samples is the spatial resolution of the observations. At the redshift of the SMACS0723 cluster ($z$=0.39), and using  a cosmology such that $\Omega_M$ = 0.3, $\Omega_{\Lambda}$ = 0.7, and a Hubble parameter H$_0$ = 70 km s$^{-1}$ Mpc$^{-1}$, the physical scale is 5.290 kpc/". Therefore assuming a NIRCam F200W PSF FWHM $\sim$ 0.066 arcsec, our spatial resolution will be 370 pc. Previous studies on the detectability of bars as a function of the spatial resolution have found a limiting resolution of $\sim$2 $\times$ PSF FWHM for a robust bar detection \citep{aguerri09, erwin18}, so we should be able to detect bars $\geq$ 740~pc in size in the NIRCam observations of SMACS0723.

It is worth mentioning that the galaxy images used for both the Virgo and Coma samples (SDSS and HST-ACS, respectively) have a spatial resolution corresponding to $\sim$ 75~pc \citep[see][for details]{mendezabreu12} at the corresponding distance of Virgo and Coma, so they would allow us to resolve bars down to sizes $\sim$ 150~pc. This difference in spatial resolution with respect to SMACS0723 might have an impact in the low-luminosity and low-mass end of the bar distribution since smaller bars are hosted in smaller galaxies \citep{aguerri09}. However, \citet{erwin18} has recently shown using a sample of galaxies from the Spitzer Survey of Stellar Structure in Galaxies \citep[S$^4$G;][]{sheth10} that only 0.02\% and 0.3\% of their bars have sizes smaller than 370 pc and 740 pc, respectively. Therefore, we expect a minimal effect on our ability to detect bars in SMACS0723 due to the spatial resolution of the observations. We note here that we are not seeking to detect inner/nuclear bars in the sample, since these can have sizes as short as 11\% of the main (outer) bar \citep{delorenzocaceres20}.

Finally, another possibility for missing bars in our SMACS0723 sample, with respect to Virgo and Coma, is the fact that bars at high redshift are expected to be shorter, since they should grow in size over time \citep{debattistasellwood00,martinezvalpuesta07}. Simulations predict that this growth in size can be significant (50\%-100\%), but it is not yet clear which galaxy parameters control the growth rate of the bars.

\section{Discussion}
\label{sec:discussion}
\subsection{Bar fraction evolution with redshift}
\label{sec:redshift}
\begin{figure}
   \centering
   \includegraphics[bb= 0 0 470 350,width=0.49\textwidth]{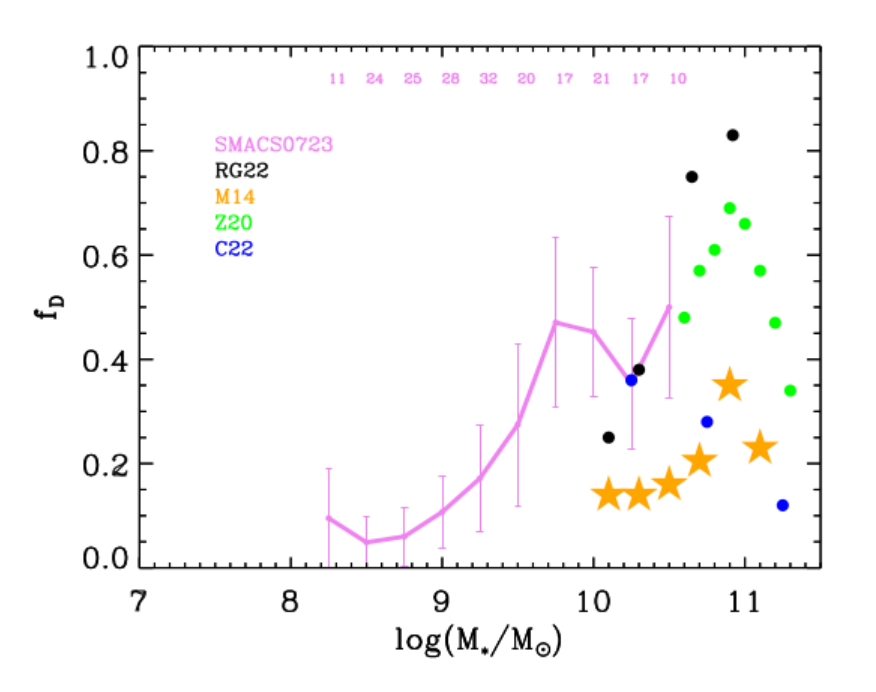}
      \caption{Comparison of the bar fraction distribution as a function of the  stellar mass of the SMACS0723 cluster (violet) with different theoretical and observational studies. The numerical simulations by \citet{rosasguevara22} at $z=0.5$, \citet{cavanagh22} at $z = 0.4$, and \citet{zhao20} are shown as black, blue, and green circles, respectively. The observational results by \citet{melvin14} at 0.4 $\leq$ $z$ $\leq$ 0.6 are shown with orange stars.
              }
         \label{fig:barfraccomp}
   \end{figure}

The evolution of the bar fraction with cosmic time has been a matter of several studies due to its implications on the settlement of the first rotationally dominated discs. As numerical simulations predict, bars can be formed spontaneously in cold discs. Since bars develop in a relatively quick phase \citep[$\leq$1 Gyr;][]{sellwood14}, and assuming that they are long lived, the presence of a bar can be used as a clock to time the formation of discs. Observationally, the studies carried out using the HST suggest a decrease of the bar fraction towards higher redshifts \citep{sheth08,cameron10}. However, the strength of this trend, as well as its dependence on galaxy properties, bar characteristics, and observational effects, is still not clear \citep{melvin14, simmons14, erwin18}. The theoretical perspective is not much different. Earlier studies based on zoom-in numerical simulations showed a clear decrease on the bar fraction with redshift \citep{kraljic12}. However, recent analyses using IllustrisTNG cosmological simulations have shown this trend might be milder when considering similar massive discs at different redshift \citep{zhao20} or even inverted \citep[higher bar fractions at higher redshifts;][]{rosasguevara22}.
Figure~\ref{fig:barfraccomp} shows the comparison of the bar fraction derived for the disc galaxies in the SMACS0723 cluster with the state of the art observational and theoretical studies. Despite performing quantitative comparisons is not straightforward due to technical and different sample selection biases, some interesting trends can be seen. From the observational side, it is clear that the measurements carried out by \citet{melvin14} in the redshift range 0.4 $\leq$ $z$ $\leq$ 0.6 provide a much lower bar fraction at all masses. One possible explanation for this difference could be the different sample selection used in \citet{melvin14}, mainly field galaxies, with respect to this work, a massive cluster. However, we go on to demonstrate in Sect.~\ref{sec:environment}, that this does not seem to be the case. Thus, we suggest that the improved capabilities of JWST with respect to HST in terms of both spatial resolution and image depth are responsible for our higher bar fraction. The comparison with numerical simulations shows in general a different trend, with simulations predicting a larger bar fraction than observations at this redshift range ($z\sim$0.4). The comparison with the IllustrisTNG50 \citep{pillepich19,nelson19} analysis by \citet{rosasguevara22} shows a similar bar fraction in their lower mass bins ($\log(M_{\star}/M_{\sun}) <$ 10.5), but it continuously grow at high masses ($\log(M_{\star}/M_{\sun}) >$ 10.5) reaching values as high as f$_{D} \sim$80\%. This high-mass end is not covered by our SMACS sample (we do not observe such massive discs) and therefore a direct comparison cannot be done. The study by \citet{zhao20} using the IllustrisTNG100 simulations obtained lower bar fractions than \citet{rosasguevara22} and, therefore, are more similar to observational results, but still higher. In general, it seems that numerical simulations are able to identify very massive disc galaxies ($\log(M_{\star}/M_{\sun}) >$ 10.5) which are not present in the clusters (neither in SMACS0723, Coma, nor Virgo) and that contain a higher fraction of bars with respect to the observations. A possible explanation of this difference might be the selection criteria of disc galaxies. Observational studies (including ours) consider a morphology-based classification between disc galaxies and elliptical, whereas most numerical simulations generally define discs with a certain threshold on the angular momentum of the particles. This is the case of the IllustrisTNG results. A different approach was followed in the analysis of the EAGLE simulations by \citet{cavanagh22}, they used a machine learning approach to morphologically classify galaxies, mimicking the way it is done in observations. The work by \citet{cavanagh22} provides overall lower bar fractions and a declining trend towards massive galaxies, thus providing a better match to observational.

\subsection{Bar fraction vs. environment}
\label{sec:environment}
\begin{figure}
   \centering
   \includegraphics[bb= 0 0 470 350,width=0.49\textwidth]{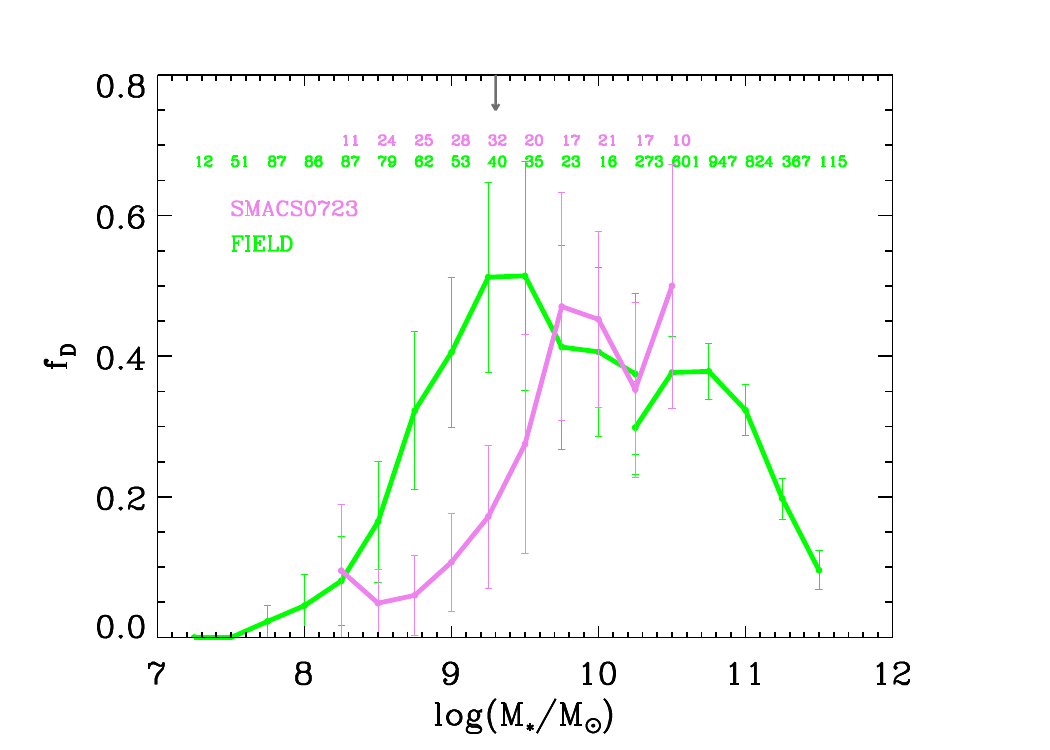}
      \caption{Comparison of the bar fraction distribution with stellar mass of the SMACS0723 cluster (violet) and in the field at $z$=0 \citep{mendezabreu12}, i.e., low-density environments (green). The arrow at ($\log(M_{\star}/M_{\sun}) \sim$ 9.3) marks the minimum thickness of discs for galaxies in the field \citep[see][]{sanchezjanssen10}}
         \label{fig:barfracenv}
   \end{figure}

Figure~\ref{fig:barfrac} shows the comparison of the SMACS0723 cluster located at $z$=0.39 with respect to the Virgo and Coma clusters located at $z$=0.0044 and $z$=0.023, respectively. As discussed in Sect.~\ref{sec:fraction}, at low luminosities/masses (M$_r \geq -18.5$~mag; $\log(M_{\star}/M_{\sun}) \leq$ 9.5) the bar fraction in the different clusters is essentially the same whereas at high luminosities/masses (M$_r < -20$~mag;  $\log(M_{\star}/M_{\sun}) >$ 10.25) the discs based (f$_{D}$) bar fractions of SMACS0723 is slightly lower than in Coma, trend that is enhanced when considering all galaxies (f$_{T}$), where the bar fraction of SMACS0723 is smaller than Virgo, with Coma showing the highest values.
Figure~\ref{fig:barfracenv} shows the bar fraction (f$_{D}$) as a function of the stellar mass for the SMACS0723 cluster and the sample of field galaxies described in \citet{mendezabreu12}. This field galaxy sample includes the galaxies analysed in \citet{aguerri09}, selected from the SDSS-DR5 \citep{adelman06} in the redshift range $0.01 < z < 0.04$, and a sample of fainter field galaxies containing all the galaxies in the SDSS-DR7 \citep{abazajian09} within $2500 < cz < 3000$ km s$^{-1}$. 
Figure~\ref{fig:barfracenv} shows a remarkable lack of bars in the low-mass regime ($\log(M_{\star}/M_{\sun}) <$ 9.75) of the SMACS0723 cluster (and therefore in Virgo and Coma) with respect to the field. The bar fraction in the field peaks at ($\log(M_{\star}/M_{\sun}) \sim$ 9.4) which roughly coincides with the minimum thickness of discs for galaxies in the field \citep[see][]{sanchezjanssen10}. The field bar fraction at its peak is $\sim$ 52\% whereas at the same mass ($\log(M_{\star}/M_{\sun}) \sim$ 9.4) the bar fraction in the SMACS0723 cluster is $\sim$ 22\%. This clearly indicate a strong influence of the environment in the low-mass discs of cluster galaxies already at $z=0.39$. The combined information that the bar fractions of SMACS0723, Virgo, and Coma clusters are the same at these galaxy masses, but different from the field at the same time, indicates that the mechanism inhibiting the formation of bars in cluster environment must be acting in early phases of the cluster assembly. At high masses ($\log(M_{\star}/M_{\sun}) >$ 10.25), Fig.~\ref{fig:barfracenv} shows that the bar fraction (f$_D$) in SMACS0723 is only marginally larger than in the local field. This reinforces the idea that previous works on field galaxies at $z\sim0.4$ are hindered by observational biases. The combined fact that SMACS0723 and Coma have comparable cluster masses, and the bar fractions among the local field, SMACS0723 (z=0.39), and Coma show a slight increase points towards an scenario with only a mild evolution in the bar fraction of high-mass galaxies during the last $\sim 4 $Gyr of evolution.

This mass-dependent influence of the environment in the bar fraction can be explained by a scenario in which interactions affect differently the structure of masive and faint discs. On the massive side, we suggest that these discs are stable enough against (tidal or galactic) interactions to keep their cold structure, thus, the increasingly larger fraction of barred galaxies in cluster with time might be explained as interactions triggering bar formation \citep{lokas16, martinezvalpuesta17, lokas20} . For faint galaxies, the same interaction due to the cluster environment might have a more destructive role. We speculate that interactions in these systems become strong enough to heat up (or destroy) the discs, thereby inhibiting bar formation and producing a lower bar fraction, with respect to the field observed in Fig.~\ref{fig:barfracenv}.

\section{Conclusions}
\label{sec:conclusions}

In this work, we study the bar fraction distribution in the SMACS0723 galaxy cluster using JWST ERO observations with the NIRCam instrument. This is the first statistical analysis of the barred population of galaxies using JWST data and it demonstrates the unique capabilities of JWST/NIRCam imaging for this kind of studies at high redshift.

We find that the bar fraction distribution in SMACS0723 is a strong function of galaxy mass, as previously shown for low redshift clusters and field galaxies \citep{mendezabreu10,mendezabreu12,erwin18}. The comparison with both Virgo and Coma clusters show that, at low luminosities and masses (M$_r \geq -18.5$~mag; $\log(M_{\star}/M_{\sun}) \leq$ 9.5), the bar fraction distribution is similar in the three cases. At high luminosities (M$_r < -20$~mag;  $\log(M_{\star}/M_{\sun}) >$ 10.25), the bar fraction distribution (computed using only disc galaxies; f$_{D}$) of SMACS0723 is only marginally lower than in Coma, with this trend getting stronger when using the overall bar fraction (computed using all cluster member galaxies; f$_{T}$). We suggest this is due to a different relative fraction of ellipticals and discs at these luminosities/masses between the clusters. We demonstrate that our results are not dependent on neither the spatial coverage of the observations for the different clusters nor on the spatial resolution of our JWST/NIRCam observations at the distance of SMACS0723 ($z=0.39$). 

We compared our results with state of the art observational and theoretical studies on the bar fraction. Numerical simulations only cover the high-mass end ($\log(M_{\star}/M_{\sun}) >$ 10.25) of the galaxy distribution and generally show larger bar fractions with respect to observations. We suggest this is due to the different selection criteria used in simulations (based on angular momentum) with respect to observations (based on morphology). At these high galaxy masses, we find a much larger bar fraction in SMACS0723 than previous works on field galaxies at the same redshift  \citep[$z\sim$ 0.4][]{melvin14}. Nevertheless, the difference is only marginal when we compare SMACS0723 with a sample of well-resolved local field galaxies \citep{mendezabreu12}. Thus, we suggest that the improved capabilities of JWST with respect to HST in terms of both spatial resolution and image depth are responsible for our higher bar fraction.

The comparison between the SMACS0723 bar fraction and that of field galaxies at $z=0$ remarks the influence of environment on the formation of bars. We find a strong drop in the bar fraction distribution of SMACS0723 low-mass galaxies (M$_r \geq -18.5$~mag; $\log(M_{\star}/M_{\sun}) \leq$ 9.75) with respect to local field galaxies. This behaviour is also found when using local clusters (Virgo and Coma), thus indicating that the mechanism inhibiting the formation of bars in cluster must acting relatively quickly after the galaxy enters into the cluster potential. On the other hand, at high luminosities and masses (M$_r < -20$~mag;  $\log(M_{\star}/M_{\sun}) >$ 10.25), the bar fraction in SMACS0723 is slightly higher than for local ($z=0$) field galaxies. This points towards a more weak influence of the environment in triggering the formation of bars at these luminosities and masses.

Our results support a scenario where cluster environment affects the formation of bars in a mass-dependent way. At high masses, the mild increase in the bar fraction of local clusters (Coma) with respect to both SMACS0723 and local field galaxies suggest a weak effect of cluster environment possibly triggering bar formation. On the other hand, low-mass galaxies show the same bar fraction in the three clusters (different redshifts) and a significant drop with respect to field galaxies at $z$ = 0, therefore suggesting
that: i) the bar fraction of low-mass galaxies in clusters is not evolving during the last $\sim$4 Gyr and ii) bar formation is severely inhibited in low-mass galaxies living in clusters.

The work presented in this paper is the first step towards a better characterization of the bar fraction (and their properties) as a function of redshift and environment. The error bars computed on the bar fraction of individual cluster are difficult to narrow down, mainly due to the fixed or otherwise limited number of cluster members. Therefore, similar analyses on a statistical number of clusters is necessary to confirm our mass-dependent scenario. Similarly, a better characterization of the bar fraction in field galaxies at different redshifts with the new JWST/NIRCam capabilities is necessary to further understand the effect of environment on the formation of bars.

\begin{acknowledgements}
     J.M.A. acknowledges the support of the Viera y Clavijo Senior program funded by ACIISI and ULL. J.M.A. acknowledges support from the Agencia Estatal de Investigaci\'on del Ministerio de Ciencia e Innovaci\'on (MCIN/AEI/ 10.13039/501100011033) under grant (PID2021-128131NB-I00) and the European Regional Development Fund (ERDF) "A way of making europe". LC would like to thank P. G. P\'erez-Gonz\'alez for the expertise acquired dealing with JWST observations. L.C acknowledges support from Agencia Estatal de Investigaci\'on del Ministerio de Ciencia e Innovaci\'on (MCIN/AEI/ 10.13039/501100011033) under grant (PGC2018-093499-B-I00) and by "European Union NextGenerationEU/PRTR". LC acknowledges financial support from Comunidad de Madrid under Atracci\'on de Talento grant 2018-T2/TIC-11612.\\
     This publication uses data generated via the Zooniverse.org platform, development of which is funded by generous support, including a Global Impact Award from Google, and by a grant from the Alfred P. Sloan Foundation
\end{acknowledgements}

%
%
\bibliographystyle{aa} 
\bibliography{reference.bib} 

\end{document}